\documentclass[conference]{IEEEtran}
\IEEEoverridecommandlockouts
\usepackage{cite}
\usepackage{amsmath,amssymb,amsfonts}
\usepackage{algorithmic}
\usepackage{graphicx}
\usepackage{textcomp}
\usepackage{xcolor}
\usepackage{algorithm}
\usepackage{algorithmic}
\usepackage{setspace}

\usepackage{amsmath}
\usepackage{epsfig}
\usepackage{graphicx}
\usepackage{times}
\usepackage{CJK}

\usepackage{makecell}
\usepackage{epsf}
\usepackage{bm} %��ѧ��ʽ�����ּӴ�
\usepackage{slashbox}
\usepackage{amsthm}
\usepackage{cite}
\usepackage{booktabs}
\usepackage{color}
\usepackage{float}
\usepackage{subfigure}

\usepackage{graphicx}

\def\BibTeX{{\rm B\kern-.05em{\sc i\kern-.025em b}\kern-.08em
    T\kern-.1667em\lower.7ex\hbox{E}\kern-.125emX}}
\begin{document}

\title{Clustering-Based Codebook Design for MIMO Communication System
\\
{\footnotesize }
%\thanks{This paper was supported by National Science and Tech-nology Major Project (2016ZX03001016), Innovation Team Project of Shaanxi Province (2017KCT-30- 02) and National Natural Science Foundation of China 6187012068.}
\thanks{This paper was supported by National Natural Science40 Foundation of China under Grant 61871321, National Science41 and Technology Major Project under Grant 2016ZX03001016,42 and Innovation Team Project of Shaanxi Province under Grant43 2017KCT-30-02.}
}
\author{\IEEEauthorblockN{ Jing Jiang\IEEEauthorrefmark{1}, Xiaojing Wang\IEEEauthorrefmark{1}, Guftaar Ahmad Sardar Sidhu\IEEEauthorrefmark{2}, Li Zhen\IEEEauthorrefmark{1}, Runchen Gao\IEEEauthorrefmark{1}}
\IEEEauthorblockA{\IEEEauthorrefmark{1}\textit{Shaanxi Key Laboratory of Information Communication Network and Security} \\
\textit{\IEEEauthorrefmark{1}Xi'an University of Posts and Telecommunications,\IEEEauthorrefmark{2}COMSATS Institute of Information Technology}\\
\IEEEauthorrefmark{1}Xi'an, China,\IEEEauthorrefmark{2}Islamabad, Pakistan \\
jiangjing@xupt.edu.cn, xiaojingwang@stu.xupt.edu.cn, guftaarahmad@comsats.edu.pk, lzhen@xupt.edu.cn,\\ runchengao@stu.xupt.edu.cn}
\and
}

\maketitle

\begin{abstract}

Codebook design is one of the core technologies in limited feedback multi-input multi-output (MIMO) communication systems. However, the conventional codebook designs usually assume MIMO vectors are uniformly distributed or isotropic. Motivated by the excellent classfication and analysis ability of clustering algorithms, we propose a K-means clustering based codebook design. First, large amounts of channel state information (CSI) is stored as the input data of the clustering, and finally divided into ${N}$  clusters according to the minimal distance. The clustering centroids are used as the statistic channel information of the codebook construction which the sum distance is minimal to the real channel information. Simulation results consist with theoretical analysis in terms of the achievable rate, and demonstrate that the proposed codebook design outperforms conventional schemes, especially in the non-uniform distribution of channel scenarios.
\end{abstract}

\begin{IEEEkeywords}
multiple-input multiple-output (MIMO), channel state inormation feedback,  codebook design, clustering
\end{IEEEkeywords}

\section{Introduction}
Multiple-input multiple-output (MIMO) technology is one of the key technologies in 4G/5G wireless communication systems, which can obviously increase the link quality and system capacity \cite{b1, b2}. The performance gains achieved by MIMO technologies depend on the perfect channel state information (CSI) at both the transmitter and the receiver. However perfect CSI is only a reasonable assumption, the limited feedback MIMO communication systems are feasible in most practical situations. For the limited feedback systems, codebook is a common solution which the receiver only sends the index of the best MIMO weighting vector from a predetermined codebook to the transmitter \cite{b3, b4}.

During the past decades, some research has been directed toward the development of codebook. Grassmannian quantization codebook was derived to maximize the minimum distance among MIMO codebook vectors. It was assumed that the channel was independently and identically distributed (\emph{i.i.d.}) and its dominant right singular vectors were uniformly distributed in the space \cite{b5, b6}. Another approach known as random vector quantization (RVQ), which assumed that MIMO vectors were isotropic and \emph{i.i.d.}, randomly generated the codebook \cite{b7, b8}. Further, discrete fourier transform (DFT) codebook scheme uniformly divided the total angular domain into ${{2^B}}$ quantized parts and generated the codebook only based on the quantization of the angle of arrive/departure \cite{b9, b10}, where ${B}$ is the number of feedback bits. One common assumption adopted in \cite{b5, b6, b7, b8, b9, b10} is that the MIMO channel is uniformly distributed or isotropic. However, in the real wireless environment, this assumption may not be always satisfied. Such as the MIMO channel was sparse in the angular domain \cite{b11} that their multi-paths are concentrated in a certain angel range. Motivated by this, a non-uniform codebook design was studied in \cite{b12} which outperformed the uniform codebook scheme. Hence, the codebook design should adapt to the arbitrary propagation environment, whether it is uniform distribution or not.

With more and more machine learning approaches being introduced to the design of wireless physical layer \cite{b13, b14, b15}, a deep learning based CSI feedback method was proposed for Massive MIMO systems, in which the convolutional neural networks transforms the channel matrices to compress representations in the receiver and proceeds the inverse transformation in the transmitter \cite{b16, b17}. Similar to \cite{b12}, \cite{b16, b17} assumed that the channel was sparse, which is also not a general hypothesis, e.g., it is not suitable for the rich scattering environment.

K-means algorithm is a famous and commonly used clustering method. Due to the classification and analysis ability for large amounts of data, it has been used in MIMO detections \cite{b18}, NOMA user clustering \cite{b19} and MIMO detection \cite{b20}. It is worth noting that the channel information could be clustered into multiple groups and each group of the channel information can be represented by the clustering centroids. Based on this, the codebook design based on K-means clustering is proposed in this paper. The major contributions in this paper are summarized as follows:

Large amounts of CSIs are divided into ${N = {2^B}}$ clusters and characteristics extracted from the clustering centroids are used as the key channel information for DFT codebook construction. The theoretical analysis proves that the sum distance of CSI to the centroid is less than the uniform vectors of the traditional codebook designs. Simulation results demonstrate that the proposed codebook design outperforms the conventional DFT schemes and can recognize and adapt to arbitrary propagation environment.

\section{System Model}
\subsection{System Model}
Consider a single cell MIMO system, ${{N_t}}$/${{N_r}}$ antennas are equipped at the transmitter/receiver, respectively. The received signal ${{\bf{y}}}$ can be described as
\begin{equation}
\bf{y = Hws + n},
\end{equation}
where ${{\bf{H}}}$ is the ${{N_r} \times {N_t}}$   MIMO channel; $\bf{s}$ denotes the
$1 \times 1$ vector of transmitted symbol; ${\bf{n}} \sim CN(0,{\sigma ^2}{\bf{I}})$ is the additive white Gaussian noise; $\bf{w}$ is the ${{N_r} \times 1}$ precoding vector. Considering the power constraint of the transmitter, we take ${{\left\| \bf{w} \right\|^2} = 1}$.

The codebook ${{\bf{C}} = [{{\bf{c}}_1},{{\bf{c}}_2},{{\bf{c}}_3},...,{{\bf{c}}_{{2^B}}}]}$ is shared by both the transmitter and the receiver, as shown in Fig. 1. The transmitter transmits CSI measurement reference signals, and the receiver selects a precoding vector ${{\bf{w}}}$ among ${N = {2^B}}$ codewords according to the following selection criteria
\begin{equation}
{{\bf{w}}} = \arg \mathop {\max }\limits_{{{\bf{c}}_i} \in {\bf{C}}} \left( {{{\left| {{\bf{H}}{{\bf{c}}_i}} \right|}^2}} \right),\forall i.
\end{equation}

With the limited feedback link, the receiver feedbacks the corresponding codeword index to the transmitter. Based on the index, the transmitter chooses the precoding matrix ${{\bf{w}}}$ from the codebook ${{\bf{C}}}$ and processes the transmitted signals with the precoding matrix ${{\bf{w}}}$.
\begin{figure}[h]
\centerline{\includegraphics[scale=0.3]{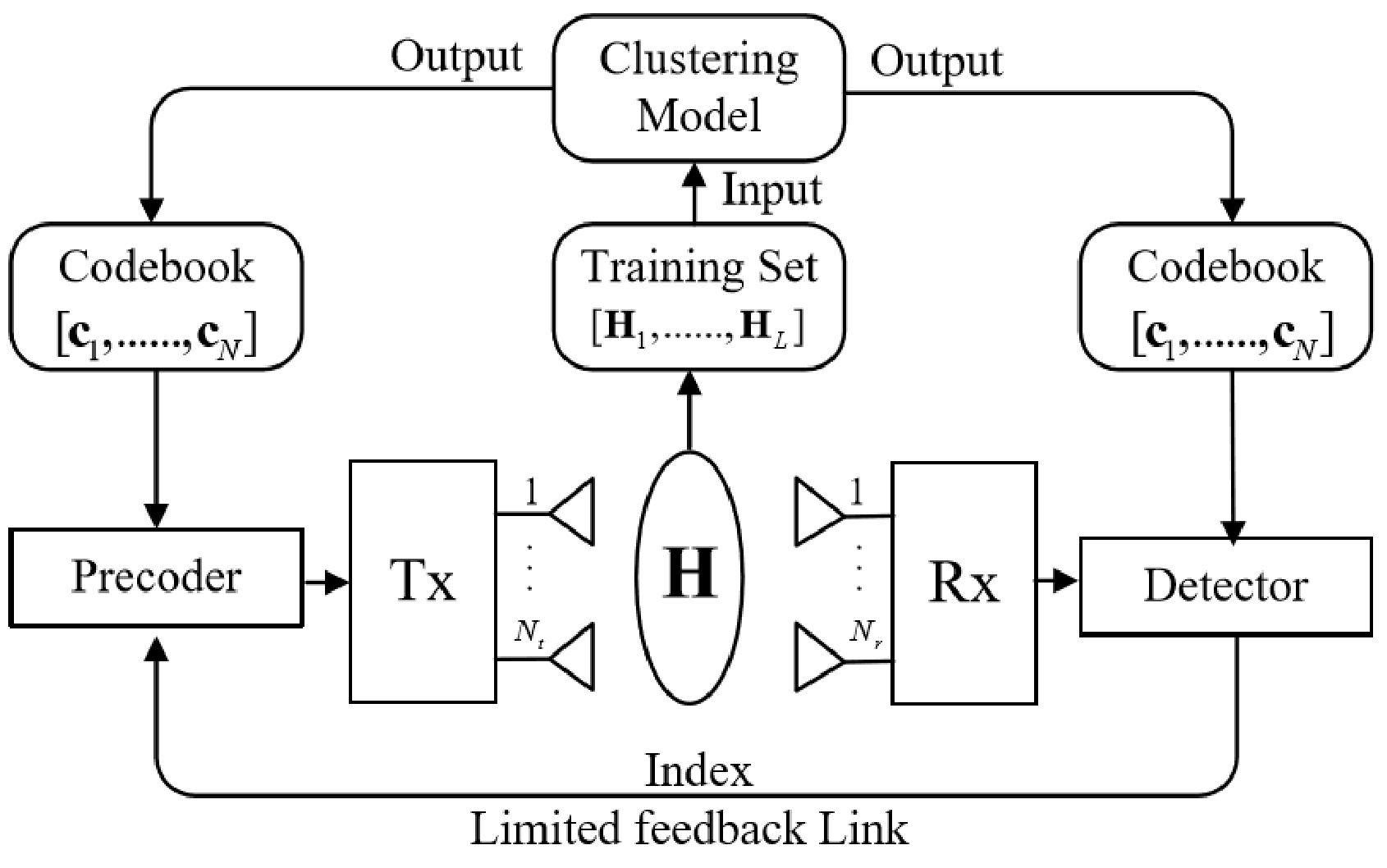}}
\caption{ Block diagram of the limited feedback MIMO systems.}
\label{fig}
\end{figure}
\subsection{Channel Model}
The matrix channel ${{\bf{H}}}$ is adopt the classical narrowband ray-based channel model and a sum of the contributions of ${p}$ propagation paths [6]. Therefore, the channel matrix ${{\bf{H}}}$ can be expressed as

\begin{equation}
{\bf{H}} = \sqrt {{N_r} \times {N_t}} \sum\limits_{i = 1}^p{{g_i}{\bf{a}}({N_r},{\phi _i}){\bf{a}}{{({N_t},{\varphi _i})}^H}},
\end{equation}
where ${{g_i}}$ is the complex gain of the ${i}$-th path, which is \emph{i.i.d.} with zero mean and unit variance; $\bf{a}( \cdot )$ is the steering vector function; ${{\phi _i}}$ and ${{\varphi _i}}$ are the cosine  angle of arrival (AoA) and angle of departure (AoD) of the ${i}$-th path, respectively. It can be expressed as\\
\[{\bf{a}}({N_t},{\varphi _i}) = \frac{1}{{\sqrt {{N_t}} }}{[1,{e^{j2\pi \frac{d}{\lambda }\cos {\varphi _i}}},...,{e^{j2\pi \frac{d}{\lambda }({N_t} - 1)\cos {\varphi _i}}}]^T}\]
\[{\bf{a}}({N_r},{\phi _i}) = \frac{1}{{\sqrt {{N_r}} }}{[1,{e^{j2\pi \frac{d}{\lambda }\cos {\phi _i}}},...,{e^{j2\pi \frac{d}{\lambda }\left( {{N_r} - 1} \right)\cos {\phi _i}}}]^T}\]
where $\lambda $ is the signal wave length, and $d$ is the spacing distance between antenna elements.

\subsection{Problem Description}

Assume that the feedback overhead ${N{\rm{ = }}{2^B}}$ is unchanged, the main goal of this paper is to improve the achievable rate in limited feedback MIMO system. Based on the channel model in Eq. (3), the achievable rate is
\begin{equation}
R = {\log _2}(1 + SNR{\left| {{\bf{Hw}}} \right|^2}),
\end{equation}
where the precoding matrix ${{\bf{w}}}$ is selected form a codebook ${{\bf{C}}}$. So ${{\bf{w}}}$ could also be denoted as ${{{\bf{c}}_i}}$, Eq. (4) is rewritten as
\begin{equation}
R = {\log _2}(1 + SNR{\left| {{\bf{H}}{{\bf{c}}_i}} \right|^2}),\forall {{\bf{c}}_i} \in {\bf{C}}.
\end{equation}

In Eq. (2), the achievable rate are related to ${{\bf{c}}_i} \in {\bf{C}}$. Hence, a codebook be designed such that the achievable rate of any user, any time and any position could be enhanced by the selected codeword in the certain scenario. In other words, the expectation of the achievable rate could be maximized. Thus, it can be formulated as follows
\begin{equation}
\mathop {\max }\limits_{\bf{C}} E[R] = \mathop {\max }\limits_{\bf{C}} E[{\log _2}(1 + SNR{\left| {{\bf{H}}{{\bf{c}}_i}} \right|^2})]
\end{equation}
\[s.t.\forall {{\bf{c}}_i} \in {\bf{C}},\;\;{\left\| {{{\bf{c}}_i}} \right\|^2} = 1,\]
Eq. (6) can be rewrite as
\[ \arg \mathop {\max }\limits_{{{\bf{c}}_i} \in {\bf{C}}} E[{\left| {{\bf{H}}{{\bf{c}}_i}} \right|^2}],\forall i.\]
%\begin{equation}
%\[ \arg \mathop {\max }\limits_{{{\bf{c}}_i} \in {\bf{C}}} E[{\left| {{\bf{H}}{{\bf{c}}_i}} \right|^2}],\forall i.\]
%\end{equation}
%which is aimed to maximize the similarity between the channel matrix H and the optimal codeword ${{{\bf{c}}_i}}$, and the optimum codeword ${{{\bf{c}}_i}}$ can always be selected from the codebook ${{\bf{C}}}$

It is observed that the optimal codeword ${{{\bf{c}}_i}}$ could be acquired if the codeword is close to its dominant right singular vectors of the real channel information. Closer the statistic channel information used to the codebook design is the real channel information, greater the performance gain achieved by the codebook design. Therefore, we defined ${d({{\bf{H}}},{{\cal H})}}$ as ${d({{\bf{H}}},{{{\cal H}}}) = {\left| {{{\bf{H}}} - {{{\cal H}}}} \right|}}$. Then, the design principle of the codebook ${{\bf{C}}}$ is expressed as

\begin{equation}
\mathop {\max }\limits_C E[R] = \mathop {\min }\limits_C E[{d^2}({\bf{H}},{\cal H})],
\end{equation}
where, ${{\cal H}}$ represent the statistic channel information used to the codebook design.

In sum, we will solve the optimization problem of the codebook design which minimize the sum-distance from the real information to the statistic channel information used to the codebook design.

\section{Codebook Design Scheme}

Clustering algorithm can divide data objects into several clusters that the similar objects are gathered into a cluster are and the distinct objects are assigned to different clusters. It represents lots of data objects by few clusters. Motivated by the feature of the clustering algorithm, we propose the codebook design in which the sum distance is minimal from the statistic channel information used to construct the codebook to the real channel information.

We firstly introduce the main idea of the proposed clustering codebook design. Then, the theoretical analysis of the minimal sum distance is addressed. In the end of this section, we descript the novel codebook design solution based on the clustering algorithm.

\subsection{Main Idea and System Process}\label{AA}
As discussed in Section II, the goal of the proposed codebook is to improve the achievable rate by minimizing the sum-distance from the real information to the statistic channel information used to construct the codebook. The main idea and system process are introduced in this section.

K-means algorithm is by far the most popular clustering scheme. The data objects in a same cluster are represented by its centroid, which is a weighted mean of the data objects and has the advantage of clear geometric and statistical meaning. Sparked by this, we cluster lots of channel information and output several centroids applied as the statistic channel information of the codebook construction. Specfically, the sum-distance from the input data to the clustering centroids is used as the feature space of K-means clustering. CSI with minimal distance is gathered into a cluster and different clusters have larger distance. And then the codebook is constructed based on the clustering centroids which possess the minimal sum distance to the real channel information.

The proposed clustering solution is operating in two phases:
\begin{enumerate}
  \item The first phase is clustering the channel information. The base station stores lots of the channel information, then the cloud processing unit processes the big CSI data with K-means clustering algorithm and outputs ${N}$ centroids which can be regarded as the statistic channel information used to the codebook construction. It is worth noting that the sum distance from the statistic channel information to the real channel information is minimal.
  \item The second phase is codebook constructing. Exploiting the centroids, the cloud processing unit constructs the codebook integrating MIMO signal processing principle.
\end{enumerate}

The proposed codebook is designed based on the cluster centroid that possess the minimal sum distance from the real information to the statistic channel information, therefore it can maximize the expectation of the achievable rate for any user in any location and any time. Moreover the cloud processing unit can periodically update the clustering processing and continually output the new codebook which can learn and adapt to the wireless environment.

\subsection{Theoretical Analysis}
We proved that as the centroid drifts the sum-distance will decrease. In the ${{i}}$-th cluster, we define ${ {\Delta _i}}$ as the gap of sum-distance between the ${{n}}$-th step and the ${\left( {n + 1} \right)}$-th step centroid,
\begin{equation}
{\Delta _i} = \sum {{{\bf{H}}_a}}  \in {\bf{cluste}}{{\bf{r}}_i}\left[ {{{\left\| {{\bf{t}}_i^n - {{\bf{H}}_a}} \right\|}^2} - {{\left\| {{\bf{t}}_i^{n + 1} - {{\bf{H}}_a}} \right\|}^2}} \right],
\end{equation}
where ${{\bf{t}}_i^n}$ represents the n-th step centroid selected in the ${{i}}$-th cluster,
\begin{equation}
\begin{aligned}
{\left\| {{\bf{t}}_i^n - {{\bf{H}}_a}} \right\|^2} = {\left\| {{\bf{t}}_i^n - {\bf{t}}_i^{n + 1} + {\bf{t}}_i^{n + 1} - {{\bf{H}}_a}} \right\|^2}\\
= {\left\| {{\bf{t}}_i^n - {\bf{t}}_i^{n + 1}} \right\|^2} + {\left\| {{\bf{t}}_i^{n + 1} - {{\bf{H}}_a}} \right\|^2} + \\
2{({\bf{t}}_i^n - {\bf{t}}_i^{n + 1})^T}({\bf{t}}_i^{n + 1} - {{\bf{H}}_a}).
\end{aligned}
\end{equation}
And then the equation is given by
\begin{equation}
{\Delta _i} = {\left\| {{{\bf{t}}_i^n} - {{\bf{t}}_i^{n + 1}}} \right\|^2}{\rm{ + }}2{({{\bf{t}}_i^n} - {{\bf{t}}_i^{n + 1}})^T}({{\bf{t}}_i^{n + 1}} - {{\bf{H}}_a}),
\end{equation}
\begin{equation}
{\Delta _i} \ge {\left\| {{{\bf{t}}_i^n} - {{\bf{t}}_i^{n + 1}}} \right\|^2}.
\end{equation}

Since ${{\left\| {{\bf{t}}_i^n - {\bf{t}}_i^{n + 1}} \right\|^2} \ge 0}$, we can infer ${{\Delta _i} \ge 0}$. That is, the sum-distance in ${{\bf{t}}_i^n}$ is larger than ${{\bf{t}}_i^{n + 1}}$.

The sum-distance in all clusters of the proposed codebook is as shown
\begin{equation}
%{dis _{uni}} = \sum\limits_{i = 1}^N {{\Delta _i}}.
{{{dis}} _{pro}} = \sum\limits_{i = 1}^N ({{\bf{H}}_a} - {{\bf{t}}_i})=\sum\limits_{i = 1}^N {{\bf{D}} _{pro}},{{{\bf{H}}_a}}  \in {\bf{cluste}}{{\bf{r}}_i},
\end{equation}
and we define the distance and the sum-distance of the uniform as ${{\bf{D}}_ {uni}}$ and ${{{dis}}_ {uni}}$ respectively. Since ${{\bf{D}}_ {uni} \ge {\bf{D}}_ {pro}}$, we can infer from this inequality that
\begin{equation}
{dis _{uni}} \ge {dis _{pro}}.
\end{equation}

According to Eq. (11), we can infer that the sum-distance from the input data to the centroid is minimized. Eq. (13) shown that the average sum-distance of the uniform quantization codebook is greater or equal to the proposed codebook.

\subsection{Clustering-based Codebook Design}

In this section, two phases are described in detail respectively.

\begin{itemize}
\item  Phase 1: K-means Based Clustering
\end{itemize}

Base station stores ${{L}}$ data objects of channel information  ${{\Omega _{\bf{H}}}{\rm{ = [}}{{\bf{H}}_1}{\rm{,}}{{\bf{H}}_2}{\rm{,}}...{\rm{,}}{{\bf{H}}_L}{\rm{]}}}$. To begin with, ${{N}}$ objects are chosen as the initial centroids, where ${N = {2^B}}$ is the number of clusters, it can be determined by the number of feedback bits ${{B}}$. We define the centroids set as: ${{\bf{T}} = [{{\bf{t}}_1},{{\bf{t}}_2},......,{{\bf{t}}_N}]}$, where ${{{\bf{t}}_i}}$ represent the centroid of ${{i}}$-th cluster.

The principle of K-means Based Clustering contains the nearest neighbor rule and the centroid condition. According to the nearest neighbor rule, each data object on behalf of the channel information is associated to the closest centroid. After all data objects are assigned, new centroid is selected from this cluster based on the centroid condition. Repeat the association of data objects and the refreshment of the centroid until the centroids and the objects in the every cluster keep unchanged.

The nearest neighbor rule and the centroid condition are described as following:

\emph{Nearest neighbor rule}:
Assign ${{{\bf{H}}_n}}$ to one of the clusters according to
\begin{equation}
d\left( {{{\bf{H}}_n},{{\bf{t}}_i}} \right) < d\left( {{{\bf{H}}_n},{{\bf{t}}_j}} \right){{\bf{H}}_n} \in {\Omega _{\bf{H}}},{{\bf{t}}_{\bf{i}}},{{\bf{t}}_{\bf{j}}} \in {\bf{T}},
\end{equation}
Then we include ${{{\bf{H}}_n}}$ into the ${{i}}$-th cluster.

The ${{i}}$-th clustering result is composed by:
\begin{equation}
{\bf{Cluste}}{{\bf{r}}_i} = [{{\bf{H}}_a},......,{{\bf{H}}_b}],i \in N,{{\bf{H}}_a} \in {\Omega _{\bf{H}}},
\end{equation}
where ${{\bf{Cluste}}{{\bf{r}}_i}}$ represent the ${{i}}$-th cluster, ${{{\bf{H}}_a}}$ is the element of the ${{i}}$-th cluster.

\emph{Centroid condition}: For each cluster, update the centroids as
\begin{equation}
\begin{array}{l}
{{\bf{t}}_i} = \arg \min \sum\limits_{{{\bf{H}}_n} \in {\bf{cluste}}{{\bf{r}}_i}} {{d^2}({{\bf{H}}_n},{\bf{T}})} \\
 = \arg \min \sum\limits_{{{\bf{H}}_n} \in {\bf{Cluste}}{{\bf{r}}_i}} {tr({\bf{I}} - {{\bf{T}}^H}{{\bf{H}}_n}{\bf{H}}_n^H{\bf{T}})} \\
 = \min \sum\limits_{{{\bf{H}}_n} \in {\bf{cluste}}{{\bf{r}}_i}} {tr({{\bf{T}}^H}{\bf{RT}})},
\end{array}
\end{equation}
where ${{\bf{R}}}$ is defined as ${{\bf{R}} = \sum {{{\bf{H}}_n}{\bf{H}}_n^H} ,{{\bf{H}}_n} \in {\bf{cluste}}{{\bf{r}}_i}}$.

After the end of the loop, we will get N centroids which possess the minimal sum distance to the real channel information in the same cluster. Then we will switch to the second phase.

It is worth mentioning that the clustering phase can be achieved in the background at the base station or the cloud processor, therefore would not influence the communication process of the limited feedback MIMO system.

\begin{itemize}
\item  Phase II: Codebook Constructing
\end{itemize}

According to phase I, we obtained ${{N}}$ clustering centroids  ${{\bf{T}} = [{{\bf{t}}_1},{{\bf{t}}_2},......,{{\bf{t}}_N}]}$, which possess the minimal sum distance to the real channel information.

We can generate a codebook by using the clustering centroids which is appropriate for various codebook design algorithms, e.g., the Singular Value Decomposition (SVD) or Discrete Fourier Transform (DFT). We will take DFT-based as an example to illustrate the detail process of codebook construction.

The codebook can be written as
${{{\bf{C}} = [{{\bf{c}}_1},{{\bf{c}}_2},......,{{\bf{c}}_N}]}}$,
where the entries of C is
\begin{equation}
{{\bf{c}}_i} = \frac{1}{{\sqrt {{N_t}} }}[1,{e^{ - j\frac{{2\pi }}{\lambda }d\cos {\theta _i}}},......,{e^{ - j\frac{{2\pi }}{\lambda }({N_T} - 1)d\cos {\theta _i}}}].
\end{equation}

As discussed in section II, we wish to choose ${{{\bf{c}}_i}}$ in order to maximized ${\mathop {\max }\limits_{{{\bf{c}}_i} \in {\bf{C}}} E[R] = \arg \mathop {\max }\limits_{{{\bf{c}}_i} \in {\bf{C}}} E[{\left| {{\bf{H}}{{\bf{c}}_i}} \right|^2}]}$. Substituting Eq. (12) into this expression, we find that
\begin{equation}
E(R) = E\left[{\log _2}\left(1 + SNR\frac{1}{{\sqrt {{N_t}} }}{\left| {{\bf{H}}\sum\limits_{n = 0}^{{N_t} - 1} {{e^{ - j\frac{{2\pi }}{\lambda }nd\cos {\theta _i}}}} } \right|^2}\right)\right]
\end{equation}
Thus, ${{\bf{\theta }}}$ is  an important parameter to increasing the achievable rate.

According to the clustering centroids, ${{\theta _i}}$ is equivalent with ${phase({{\bf{t}}_i})}$. The Eq. (11) could also be expressd as

${{{\bf{c}}_i} = \frac{1}{{\sqrt {{N_t}} }}[1,{e^{ - j\frac{{2\pi }}{\lambda }d\cos {{(phase({{\bf{t}}_i}))}}}},...,}$
\begin{equation}
{e^{ - j\frac{{2\pi }}{\lambda }({N_T} - 1)d\cos phase({{\bf{t}}_i})}}{]^T},
\end{equation}

\begin{equation}
{phase({{\bf{t}}_i}) = {{\log {{\bf{t}}_i}\sqrt {{N_t}} } \mathord{\left/
 {\vphantom {{\log {{\bf{t}}_i}\sqrt {{N_t}} } j}} \right.
 \kern-\nulldelimiterspace} j}},
\end{equation}
and then ${{\bf{\theta }} = [phase({{\bf{t}}_1}),phase({{\bf{t}}_2}),......,phase({{\bf{t}}_N})]}$, which is satisfies
\[\bigcap\limits_{i = 1,......,N} {phase({{\bf{t}}_i})}  = \phi. \]

Therefore, the non-uniform quantization codebook is given by
\begin{equation}
{\bf{C}}(:,i) = \frac{1}{{\sqrt {{N_t}} }}*{e^{ - j\frac{{2\pi }}{\lambda }d({N_T} - 1)\cos (phase({{\bf{t}}_i}))}}.
\end{equation}

Then we summarize the procedures in Algorithm 1.

\begin{tabular}{lcl}
\\  \toprule
$\bf{Algorithm\ 1}$:  clustering based codebook\\ \midrule
${{\bf{Input}}:{{\bf{N}}_{\rm{t}}},{\bf{B}},{\Omega _{\bf{H}}} = \{ {{\bf{H}}_{\bf{1}}},{{\bf{H}}_{\bf{2}}},......,{{\bf{H}}_{\bf{L}}}\} }$\\
${{\bf{Output}}:{\bf{C}}}$\\
\textbf{First stage}:\\
	Chooses N samples as the initial centroids, written as\\
~~~~~~~~~~~~${{\bf{T}} = [{{\bf{t}}_1},{{\bf{t}}_2},......,{{\bf{t}}_N}]}$\\
\textbf{for} ${a = 1:N}$\\
~~~~Calculate the distance from point to the centroids:\\
~~~~~~~~~~~~${d(a) = {\left\| {{{\bf{H}}_n} - {{\bf{t}}_a}} \right\|^2}}$\\
\textbf{end}\\
${{\bf{d}} = [d(1),d(2),...,d(N)]}$\\%where ${{d_a}}$ represent the distance between ${{{\bf{H}}_n}}$ to all centroids\\
${i = \mathop {\min }\limits_{i \in N} ({\bf{d}})}$ \\divide  ${{\bf{H}_n}}$ into ${{\bf{cluste}}{{\bf{r}}_i}}$ \\

	Update the centroid of each cluster according to the \\criteria\\
~~~~~~~~~~~~${{\bf{t}}_i^{opt} = \min \sum\limits_{{{\bf{H}}_n} \in {\bf{cluste}}{{\bf{r}}_i}} {tr({{\bf{T}}^H}{\bf{RT}})} ,i \in N}$\\
~~~~~~~~~~~~${{\bf{R}} = \frac{1}{{{N_t}}}{{\bf{H}}_n}{\bf{H}}_n^h}$\\
	Until the centroids remain unchanged\\
The final centroids remark as:\\
~~~~~~~~~~~~${{\bf{T}} = [{{\bf{t}}_1},{{\bf{t}}_2},......,{{\bf{t}}_N}]}$\\
	\textbf{Second stage:}\\
Calculate the angle information of ${phase({{\bf{t}}_i})}$ contained \\in ${{{\bf{t}}_i}}$\\
~~~~~~~~~~~~${phase({{\bf{t}}_i}) = {{\log {{\bf{t}}_i}\sqrt {{N_t}} } \mathord{\left/
 {\vphantom {{\log {{\bf{t}}_i}\sqrt {{N_t}} } j}} \right.
 \kern-\nulldelimiterspace} j}}$\\
~~~~~~~~~~~~${{\bf{\theta }} = [phase({{\bf{t}}_1}),phase({{\bf{t}}_2}),......,phase({{\bf{t}}_N})]}$\\
non-uniform quantization codebook can be written as:\\
~~~~~~~~~~~~${{\bf{C}}(:,i) = \frac{1}{{\sqrt {{N_t}} }}{e^{ - j\frac{{2\pi }}{\lambda }d({N_T} - 1)\cos (phase({\bf{t_i}}))}}}$
\\ \bottomrule
\end{tabular}

\section{Simulation Results}

%In this section, we evaluate the performance of the proposed codebook design algorithm by computational simulations. The clustering-based codebook design algorithm and the conventional codebook design algorithm are chosen as benchmark. Without loss of generality, the parameters for MIMO limited feedback system are the same as those in \cite{b22, b23}. For ${{{p}}}$ multi-paths, AoD is set to be ${{\bf{\theta }} ={{\bf{\theta}}_0}} + {\Delta\theta} $, where   ${{\theta _0}}$ is a constant that is randomly selected within ${[0,2\pi ]}$. ${\Delta \theta }$ is the angle spread value, which is randomly selected within ${ [0,{\theta _{BW}}]}$, where ${{\theta _{BW}}}$ is the angle spread parameter to distinguish the different channel condition. In the simulations, ${{\theta _{BW}}}$ is set to be ${{\pi  \mathord{\left/{\vphantom {\pi  6}} \right.\kern-\nulldelimiterspace} 6}}$, ${{\pi  \mathord{\left/{\vphantom {\pi  3}} \right.\kern-\nulldelimiterspace} 3}}$, and ${\pi }$, corresponding to the nonuniform scenario with small angle spread, the non-uniform scenario with middle angle spread, and the uniform scenario, respectively.

In this section, the performance of our proposed codebook design is evaluated using computer simulations. For comparison, the clustering-based codebook design and the conventional codebook design are also simulated with various channel condition and antenna configuration. Consider a MIMO limited feedback system as described in \cite{b22, b23}. Both the transmitter and the receiver configure the uniform linear array (ULA) with 0.5$\lambda $ antenna spacing. For p multi-paths in Eq. (2), the mean angles of AoD are set to be ${{\bf{\varphi }} ={{\bf{\varphi}}_0}} + {\Delta\varphi} $, where ${{\varphi _0}}$ is a constant that is randomly selected within ${[0,2\pi ]}$ and the angle spreads are set as ${\Delta \varphi }$. ${\Delta \varphi }$ is randomly selected within${ [0,{\varphi _{BW}}]}$. The angle spread ${{\varphi _{BW}}}$ is the parameter distinguishing the different channel distribution. In the simulations, the angel range of the traditional DFT codebook construction is ${\pi}$. Correspondingly,
${{\varphi _{BW}}}$ is equal to  ${{\pi  \mathord{\left/{\vphantom {\pi  6}} \right.\kern-\nulldelimiterspace} 6}}$, ${{\pi  \mathord{\left/{\vphantom {\pi  3}} \right.\kern-\nulldelimiterspace} 3}}$, and ${\pi }$ which refers to the non-uniform scenario with small angle spread, the non-uniform scenario with middle angle spread, and the uniform scenario, respectively.

Fig. 2, Fig. 3 and Fig. 4 compare the achievable rate of the proposed codebook scheme with different numbers of transmitting antennas. In un-uniform scenario (${{\varphi _{BW}}}$ is equal to  ${{\pi  \mathord{\left/{\vphantom {\pi  6}} \right.\kern-\nulldelimiterspace} 6}}$, ${{\pi  \mathord{\left/{\vphantom {\pi  3}} \right.\kern-\nulldelimiterspace} 3}}$) shown in Fig. 2 and Fig. 3, the proposed codebook scheme outperforms the traditional DFT codebook scheme. More importantly, the performance gaps between the proposed schemes and the traditional codebook designs are more obvious as the angle spread decreases. It can be explained that the proposed codebook design could learn the channel statistical characteristics, but the uniform assumption of the traditional codebook designs is contradictory with the real environment. On the other hand, in uniform scenario shown in Fig. 4, the proposed algorithm achieves the same performance curve with the conventional one in uniform distribution. Because the statistical characteristics learned the proposed codebook design are same the assumptions of the traditional codebook designs, and both of them coordinate with the real channel information.

%Fig. 2, Fig. 3 and Fig. 4 depict the achievable rates versus SNR for different $N_t$ and the angle spread. First, it can be observed from Fig. 2 and Fig. 3 that the achievable rates of the proposed and benchmark schemes increase with SNR and $N_t$, which is in accordance with our analysis. On the other hand, our proposed algorithm outperforms the traditional DFT codebook scheme in un-uniform scenario and the achievable rate of traditional scheme is insensitive to the change of angle spread. In addition, Fig. 4 shows that the proposed algorithm achieves the same performance curve with the conventional one in uniform distribution, which indirectly proves the feasibility and effectiveness of the proposed algorithm.

\begin{figure}[htb]
%\centerline{\includegraphics[scale=0.50]{achi301}}
\centerline{\includegraphics[scale=0.50]{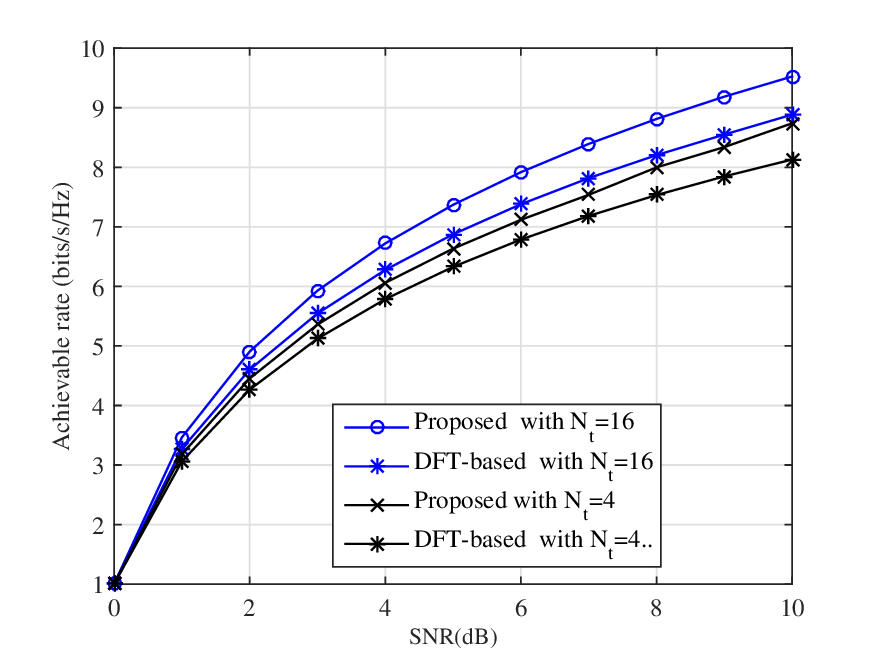}}
\caption{ Achievable rate comparison with different numbers of transmitting antennas, where ${{\varphi _{BW}}=30^\circ}$.}
\label{fig}
\end{figure}
\begin{figure}[htb]
%\centerline{\includegraphics[scale=0.50]{achi501}}
\centerline{\includegraphics[scale=0.50]{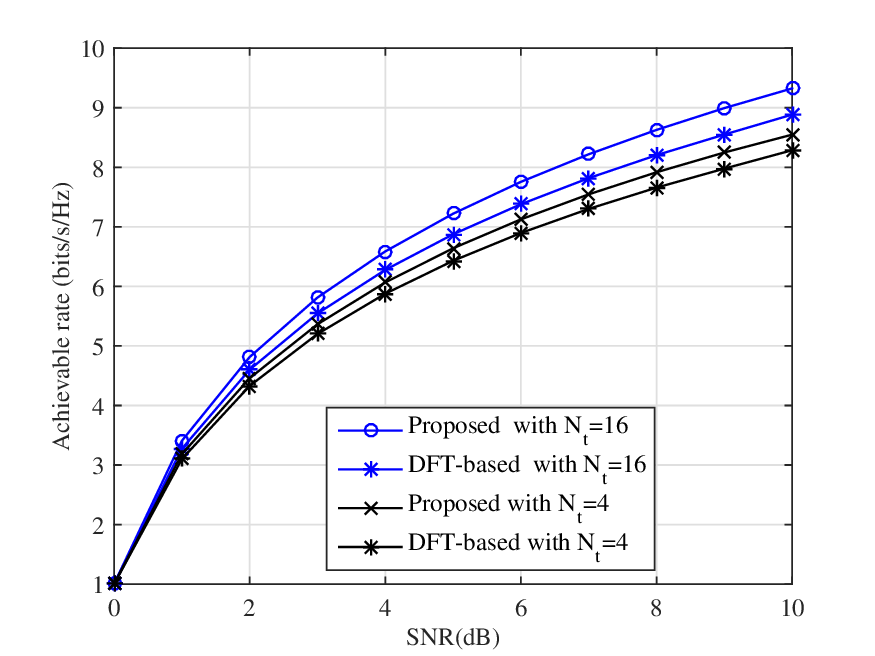}}
\caption{ Achievable rate comparison with different numbers of transmitting antennas, where ${{\varphi _{BW}}=60^\circ}$.}
\label{fig}
\end{figure}
\begin{figure}[htb]
%\centerline{\includegraphics[scale=0.50]{achi1801}}
\centerline{\includegraphics[scale=0.50]{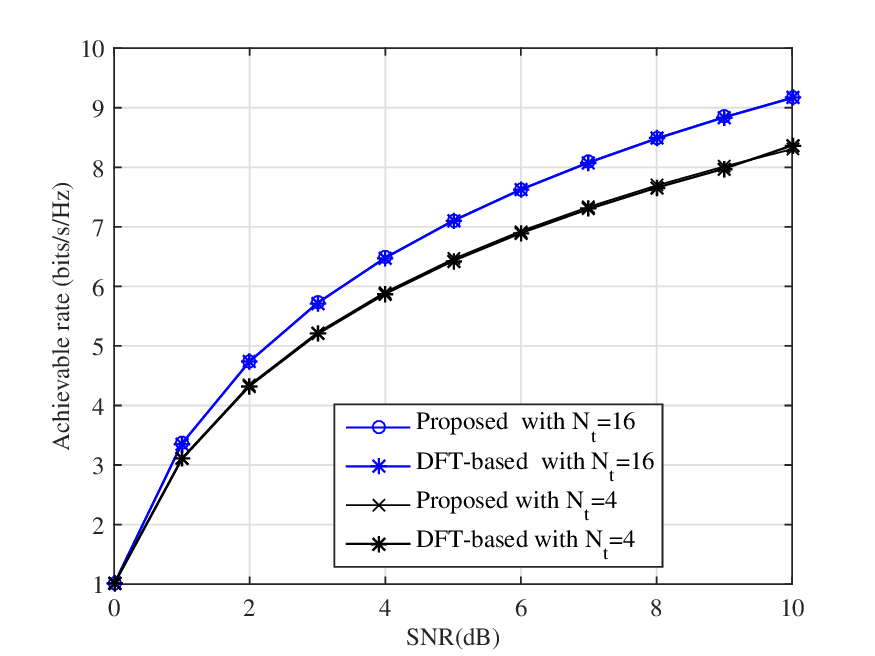}}
\caption{ Achievable rate comparison with different numbers of transmitting antennas, where ${{{{\varphi _{BW}}}} = 180^\circ }$.}
\label{fig}
\end{figure}

\section{Conclusion}
In this paper, we developed a novel MIMO codebook design based on K-means clustering algorithm. The proposed codebook design leverage the K-means clustering to form ${N}$ clustering centroids in which the sum distance is minimal from the statistic channel information used to construct the codebook to the real channel information. Therefore the proposed codebook design can maximize the expectation of the achievable rate. Both the theoretical analyses and simulation results demonstrate that the proposed codebook design outperforms conventional schemes, especially in the non-uniform distribution of channel scenarios. Fortunately, with periodically updating the clustering processing, the proposed codebook design can continually learn and adapt to the real environment.

%\section*{References}

\vspace{12pt}

\end{document}